\pdfoutput=1
\pdfoutput=1
\pdfoutput=1
\pdfoutput=1
\pdfoutput=1

\batchmode
\makeatletter
\def\input@path{{C:/Users/Achel/Desktop/13ene//}}
\makeatother
\documentclass[10pt,english]{article}
\usepackage[T1]{fontenc}
\usepackage[latin1]{inputenc}
\usepackage[letterpaper]{geometry}
\usepackage{rotfloat}
\usepackage{amsmath}
\usepackage{amssymb}
\usepackage{graphicx}
\usepackage{esint}
\usepackage[dot]{bibtopic}

\makeatletter
\usepackage{amscd}

\usepackage{amsfonts}

\usepackage{babel}\makeatother

\makeatother

\usepackage{babel}

\makeatother

\usepackage{babel}
\begin{document}

\title{\textbf{Option Pricing of Twin Assets}}

\author{Marcelo J. Villena%
\thanks{Adolfo Ib\'a\~nez University, Chile. Corresponding author email: marcelo.villena@uai.cl.%
}\quad{}and\enskip{}Axel A. Araneda%
\thanks{Adolfo Ib\'a\~nez University, Chile.%
}}
\maketitle
\begin{abstract}
How to price and hedge claims on nontraded assets are becoming increasingly
important matters in option pricing theory today. The most common
practice to deal with these issues is to use another similar or \textquoteleft{}closely
related\textquoteright{} asset or index which is traded, for hedging
purposes. Implicitly, traders assume here that the higher the correlation
between the traded and nontraded assets, the better the hedge is expected
to perform. This raises the question as to how \textquoteleft{}closely
related\textquoteright{} the assets really are. In this paper, the
concept of twin assets is introduced, focusing the discussion precisely
in what does it mean for two assets to be similar. Our findings point
to the fact that, in order to have very similar assets, for example
identical twins, high correlation measures are not enough. Specifically,
two basic criteria of similarity are pointed out: i) the coefficient
of variation of the assets and ii) the correlation between assets.
From here, a method to measure the level of similarity between assets
is proposed, and secondly, an option pricing model of twin assets
is developed. The proposed model allows us to price an option of one
nontraded asset using its twin asset, but this time knowing explicitly
what levels of errors we are facing. Finally, some numerical illustrations
show how twin assets behave depending upon their levels of similarities,
and how their potential differences will traduce in MAPE (mean absolute
percentage error) for the proposed option pricing model. 
\end{abstract}
\begin{center}
\par\end{center}{\footnotesize \par}

\newpage{}

\section{Introduction}

Usually, financial markets in the real world are not complete. Indeed,
for instance, how to price and hedge claims on nontraded assets are
becoming increasingly important matters in option pricing theory today.
The most common practice to deal with this issue is to use for hedging
purposes another similar or \textquoteleft{}closely related\textquoteright{}
asset or index which is traded. The finance literature has focused
on finding the best hedging strategy, usually using a utility maximization
approach, see for example \cite{henderson2002valuation,davis2006optimal}.
It is well known that this cross hedging mechanism in an incomplete
financial market creates what has been called basis risk, see \cite{chang2003cross,adam2011cross}. 

In this line of research, implicitly traders assume here that the
higher the correlation between the traded and nontraded assets, the
better the hedge is expected to perform. This is certainly an advantage
of this strategy today, since over the past ten years, cross-asset
correlations roughly doubled in international financial markets%
\footnote{JPMorgan developed an analysis on the correlation between 45 developed
world and emerging market country equity benchmarks contained in the
MSCI All Country World Index, see \cite{JPMorgan2011}. Its results
show that, over the past 20 years, the average correlation between
these country benchmarks roughly doubled, from 30\% to 60\%. %
}. On the other hand, there are many examples of financial problems
of this kind, with one nontraded assets, where cross hedging is needed:
for example an exporter  of a particular commodity when the hedging
instrument relates to another commodity, \cite{wong2012cross}, or
commodity-currency cross-hedges, see \cite{chen2012better}. 

Nevertheless, cross hedging in an incomplete financial market raises
the question as to what really are similar or \textquoteleft{}closely
related\textquoteright{} assets, beyond the high correlation between
the traded and nontraded assets. In fact, there are not many studies
analysing the effectiveness of this definition of assets similarity.
In this paper, we introduce the concept of twin assets, focusing the
discussion precisely in what does it mean to be similar. Our findings
point to the fact that, in order to have very similar assets, for
example identical twins following our metaphor, high correlation measures
are not enough, we need to consider in our characterization of similarity,
a normalise measure of relative volatility. Specifically, we consider
two basic criteria of similarity to define twin assets: i) the coefficient
of variation of the assets, and ii) the correlation between assets. 

The structure of the paper is as follows. In section 2, firstly, a
method to measure the level of similarity between assets is proposed,
and secondly, an option pricing model of twin assets is developed.
The proposed model allow us to price an option of one nontraded asset
using its twin asset, but this time knowing explicitly what levels
of errors we are facing. In section 3, some numerical illustrations
show how twin assets behave depending upon their levels of similarities,
and how their potential differences will traduce in MAPE (mean absolute
percentage error) for the proposed option pricing model. Finally,
some conclusions and future research are presented.

\pagebreak{}

\section{The Model}

In this section our model is presented. Firstly, we introduce two
parameters than help us to measure the level of similarity between
assets. Latter, with the definition of twin assets, an option pricing
model is developed, considering implicitly the parameters of similarities
mentioned above.

\subsection{Looking for Twin Assets}

Let $S_{i}^{t}$ be the value of the stock $i$ and $S_{j}^{t}$ the
value of the stock $j$, both at time $t$. Let us assume that they
are governed by the following stochastic differential equations:

\begin{equation}
\textrm{d}S_{i}^{t}=\mu_{i}S{}_{i}^{t}\textrm{d}t+\sigma_{i}S_{i}^{t}\textrm{d}W_{i}^{t}\label{eq:dSi}
\end{equation}

\begin{equation}
\textrm{d}S_{j}^{t}=\mu_{j}S{}_{j}^{t}\textrm{d}t+\sigma_{j}S_{j}^{t}\textrm{d}W_{j}^{t}\label{eq:dSj}
\end{equation}

\noindent where $\sigma_{i}$ and $\sigma_{j}$ are constants. $W_{i}^{t}$
and $W{}_{j}^{t}$ are Gauss-Wiener processes that maintain the following
relationship:

\begin{equation}
\textrm{d}W_{i}^{t}\cdot\textrm{d}W_{j}^{t}=\rho\textrm{d}t\label{eq:corr}
\end{equation}
\\

The last equation implies that returns of stock $i$ and $j$ are
correlated, and that the value of that correlation is $\rho$. An
equivalent expression for \eqref{eq:corr} is:

\begin{equation}
\textrm{d}W_{i}^{t}=\textrm{\ensuremath{\rho}d}W_{j}^{t}+\textrm{\ensuremath{\sqrt{1-\rho^{2}}}d}\tilde{W}^{t}\label{eq:corr-1}
\end{equation}

\noindent with $\tilde{W}^{t}$ defined as an independent Gauss-Wiener
processes.\\

Until here, we have followed the standard procedure for modelling
similar assets. Nevertheless, when two highly correlated assets are
modelled, they are not necessarily going to follow exactly the same
trajectory, neither, they are going to finish in a similar point.
The only thing we can say about them is that they are going to go
up and down at the same time, but the amount of the ups and downs
will depend upon the volatility. On the other hand, the volatility
has to be corrected by the mean (two assets with equal volatilities
but different means are not going to follow similar trajectories neither).
So we need to include both the volatility and the mean as variables
in order to define similar assets. Thus, we introduce a normalise
measure of relative volatility, an $\alpha$ parameter, defined by
quotient of the the coefficients of variation of each asset, which
is given by the following expression:

\begin{equation}
\alpha\equiv\frac{\left(c_{v}\right)_{i}}{\left(c_{v}\right)_{j}}=\frac{\sigma_{i}\mu_{j}}{\sigma_{j}\mu_{i}}
\end{equation}

Then, using the It�'s calculus, we obtain an analytical solution for
$S_{i}$ and $S_{j}$ at time $T>t'$:\\

\begin{alignat}{1}
S_{i}^{T} & =S_{i}^{t'}e^{\left(\mu_{i}-\frac{1}{2}\sigma_{i}^{2}\right)\left(T-t'\right)+\sigma_{i}\left(\rho W_{j}^{T-t'}+\sqrt{1-\rho^{2}}\tilde{W}^{T-t'}\right)}\label{eq:SiT}
\end{alignat}

\begin{eqnarray}
S_{j}^{T} & = & S_{j}^{t'}e^{\left(\mu_{j}-\frac{1}{2}\sigma_{j}^{2}\right)\left(T-t'\right)+\sigma_{j}W_{j}^{T-t'}}\label{eq:SjT}
\end{eqnarray}

\noindent where $S_{i}^{t'}$and $S_{j}^{t'}$ are given values at
the time $t'$ of the stock $i$ and $j$ respectively.\\

From \eqref{eq:SiT} and \eqref{eq:SjT}we can build expressions for
the returns of the stocks $S_{i}$ and $S_{j}$, called $R_{i}$ and
$R_{j}$ respectively, over the time span $T-t'$:

\begin{equation}
R_{i}^{T-t'}=\ln\left(\frac{S_{i}^{T}}{S_{i}^{t'}}\right)
\end{equation}

\begin{equation}
R_{j}^{T-t'}=\ln\left(\frac{S_{j}^{T}}{S_{j}^{t'}}\right)
\end{equation}

After replacing and rearranging some terms we have:\\
\begin{align}
R_{j}^{T-t'} & =\alpha\frac{\sigma_{j}}{\sigma_{i}}R_{i}^{T-t}+\frac{1}{2}\sigma_{j}\left(\alpha\sigma_{i}-\sigma_{j}\right)\left(T-t'\right)+\sigma_{j}\left(1-\rho\alpha\right)W_{j}^{T-t'}-\alpha\sigma_{j}\sqrt{1-\rho^{2}}\tilde{W}^{T-t'}\label{Rj_i}
\end{align}

\noindent and hence:
\begin{align}
S_{j}^{T}e^{\sigma_{j}\left(1-\rho\alpha\right)W_{j}^{T-t'}-\alpha\sigma_{j}\sqrt{1-\rho^{2}}\tilde{W}^{T-t'}} & =S_{j}^{t'}\left(\frac{S_{i}^{T}}{S_{i}^{t'}}\right)^{\alpha\frac{\sigma_{j}}{\sigma_{i}}}e^{\frac{1}{2}\sigma_{j}\left(\alpha\sigma_{i}-\sigma_{j}\right)\left(T-t'\right)}\label{Sj_i}
\end{align}
\\

Equations \eqref{Rj_i} and \eqref{Sj_i} give relationships between
each pair of returns and stocks respectively, allowing us to express
the stock (return) $j$ in terms of the stock (return) $i$. These
equalities, given their stochastic nature, can be modelled only when
the stochastic terms (i.e. $W_{j}^{T-t'}$ and $\tilde{W}^{T-t'}$)
are known. Therefore, we propose the following approximation: 

\begin{align}
S_{j}^{T}\thickapprox S_{j}^{t'}\left(\frac{S_{i}^{T}}{S_{i}^{t'}}\right)^{\alpha\frac{\sigma_{j}}{\sigma_{i}}}e^{\frac{1}{2}\sigma_{j}\left(\alpha\sigma_{i}-\sigma_{j}\right)\left(T-t'\right)}e^{\sigma_{j}\left(1-\rho\alpha\right)W_{x}^{T-t'}-\alpha\sigma_{j}\sqrt{1-\rho^{2}}W_{y}^{T-t'}}\label{Sj_i est}
\end{align}

\noindent where $W_{x}^{T-t}$and $W_{y}^{T-t}$ are independent
Brownian motions. \\

We see that \eqref{Sj_i est} is governed by the values of $\rho$
and $\alpha$. If simultaneously $\rho\rightarrow1$ and $\alpha\rightarrow1$
the error of the approximation will be the minimum. Thus, we will
define as \emph{twin assets,} when both $\rho$ and $\alpha$ parameters
come close to the unity (i.e. the returns of the assets are totally
correlated and have the same (similar) coefficients of variation).
Alternatively, we can have different levels of similarity depending
upon the values of $\rho$ and $\alpha$.

Finally, the expression \eqref{Sj_i est} can be written in a more
compact form as:
\begin{equation}
S_{j}^{T}\thickapprox AB\left(S_{i}^{T}\right)^{\alpha\frac{\sigma_{j}}{\sigma_{i}}}\label{Sj_i compact}
\end{equation}

\noindent being A a deterministic term and B a stochastic term given
by: 

\begin{equation}
A=S_{j}^{t'}\left(S_{i}^{t'}\right)^{-\alpha\frac{\sigma_{j}}{\sigma_{i}}}e^{\frac{1}{2}\sigma_{j}\left(\alpha\sigma_{i}-\sigma_{j}\right)\left(T-t'\right)}\label{eq:A}
\end{equation}

\begin{equation}
B=e^{\sigma_{j}\left(1-\rho\alpha\right)W_{x}^{T-t'}-\alpha\sigma_{j}\sqrt{1-\rho^{2}}W_{y}^{T-t'}}\label{eq:B}
\end{equation}
\\

From these equations we can clearly see that the correlation between
assets is a necessary but not sufficient condition in order to have
similar or twin assets. As we will see in the next sections, the predictive
performance of option pricing and hedging will depend crucially on
the $\rho$ and $\alpha$ parameters.

\subsection{On the Option Pricing of Twin Assets}

An important practical question for traders and investors in the presence
of incomplete markets is how to price and hedge what we have called
twin options. For example, in the case of a nontraded asset or a given
option, if we have a twin asset, defined by their similarities in
terms of the coefficient of variation and correlation, we could have
a potential pricing and hedging of the asset or option, but this time
with a measure of the errors involved.

According with the traditional approach, a derivative $c_{j}$ with
underlying asset $S_{j}$ follow the Black-Scholes equation \cite{black1973pricing}:

\begin{equation}
rc_{j}=\frac{\partial c_{j}}{\partial t}+\frac{1}{2}\sigma_{j}^{2}S_{j}^{2}\frac{\partial^{2}c_{j}}{\partial S_{j}^{2}}+rS_{j}\frac{\partial c_{j}}{\partial S_{j}}\label{eq:BS-j}
\end{equation}

\noindent Being $r$ the risk-free rate of interest for a term.\\

If we consider $c_{j}$ as a vanilla call option with time to maturity
$T$ (fixed the starting date of the contract in $t=0$) and Strike
Price $K_{j}$, the value of $c_{j}$ is given by the Black-Scholes
formula:

\begin{equation}
c_{j}\left(T-t\right)=S_{j}^{t}N\left(d_{1_{j}}\right)-K_{j}e^{-r\left(T-t\right)}N\left(d_{2_{j}}\right)\label{eq:BScj}
\end{equation}

\noindent where $N(\cdot)$ is the cumulative normal density function
and

\begin{equation}
d_{1_{j}}=\frac{\ln\left(\frac{S_{j}^{t}}{K_{j}}\right)+\left(r+\frac{\sigma_{j}^{2}}{2}\right)\left(T-t\right)}{\sigma_{j}\sqrt{T-t}}
\end{equation}

\begin{equation}
d_{2_{j}}=\frac{\ln\left(\frac{S_{j}^{t}}{K_{j}}\right)+\left(r-\frac{\sigma_{j}^{2}}{2}\right)\left(T-t\right)}{\sigma_{j}\sqrt{T-t}}
\end{equation}
\\

On the another hand, we can link $c_{j}$ with an option over the
stock $S_{i}$. Since $A>0$ and $B>0,$ by \eqref{Sj_i compact}
and considering the payoff function of $c_{j}$:

\begin{equation}
\left\{ S_{j}^{T}-K_{j}\right\} ^{+}\thickapprox AB\left\{ \left(S_{i}^{T}\right)^{\alpha\frac{\sigma_{j}}{\sigma_{i}}}-K_{i}\right\} ^{+}\label{eq:Sj power}
\end{equation}

\noindent where $K_{i}=K_{j}/\left(AB\right)$. \\

Now, using \eqref{eq:Sj power} the price of $c_{j}$ also is related
with the value of $S_{i}$. By risk neutral valuation, we have:

\begin{align}
e^{-r\left(T-t\right)}\mathcal{\mathbb{E}}\left[\left\{ S_{j}^{T}-K_{j}\right\} ^{+}\right]\thickapprox e^{-r\left(T-t\right)}\mathcal{\mathbb{E}}\left[AB\left\{ \left(S_{i}^{T}\right)^{\alpha\frac{\sigma_{j}}{\sigma_{i}}}-K_{i}\right\} ^{+}\right]\label{eq:Sj powerrnv}
\end{align}

The left side of \eqref{eq:Sj powerrnv} correspond to the price of
$c_{j}$ at time $T-t$ (see \eqref{eq:BScj}). Then,

\begin{equation}
c_{j}\thickapprox e^{-r\left(T-t\right)}\intop_{K_{i}^{\frac{\sigma_{i}}{\alpha\sigma_{j}}}}^{\infty}\left(\left(S_{i}^{T}\right)^{\alpha\frac{\sigma_{j}}{\sigma_{i}}}-K_{i}\right)l\left(S_{i}^{T}\right)\mathrm{d}S_{j}^{T}\label{eq:Sj int}
\end{equation}

\noindent being $l\left(S_{i}^{T}\right)$ the probability distribution
of $S_{i}^{T}$. Using \eqref{eq:SiT} and the fact that $W_{i}^{T}=w\sqrt{\left(T-t\right)}$,
with $w\thicksim f(w)$, being $f$ the normal standard distribution%
\footnote{i.e. $w$ is standard normal variable%
}; \eqref{eq:Sj int} is transformed into:

\begin{align}
c_{i}\left(T-t\right) & \thickapprox AB\frac{e^{-r\left(T-t\right)}}{\sqrt{2\pi}}\intop_{-g_{2_{i}}}^{\infty}\left[\left(S_{i}^{t}e{}^{\left(r-\frac{1}{2}\sigma_{i}^{2}\right)\left(T-t\right)+x\sigma_{i}\sqrt{T-t}}\right)^{\alpha\frac{\sigma_{j}}{\sigma_{i}}}-K_{i}\right]e^{-\frac{w^{2}}{2}}\mathrm{d}w\label{eq:pricing}
\end{align}

\noindent where,

\begin{equation}
g_{2}=\frac{\ln\left(\frac{S_{i}^{t}}{K_{i}^{\sigma_{i}/\alpha\sigma_{j}}}\right)+\left(r-\frac{1}{2}\sigma_{i}^{2}\right)\left(T-t\right)}{\sigma_{i}\sqrt{T-t}}
\end{equation}
\\

In order to solve, we separate the left side into two integrals:

\begin{equation}
c_{i}\left(T-t\right)\thickapprox I_{1}-I_{2}
\end{equation}

\noindent being,

\begin{align}
I_{1}=\frac{AB\left(S_{i}^{t}\right)^{\alpha\frac{\sigma_{j}}{\sigma_{i}}}e^{-r\left(T-t\right)}}{\sqrt{2\pi}}\intop_{-g_{2_{i}}}^{\infty}e{}^{\left[\left(r-\frac{1}{2}\sigma_{i}^{2}\right)\left(T-t\right)+x\sigma_{i}\sqrt{T-t}\right]\alpha\frac{\sigma_{j}}{\sigma_{i}}}e^{-\frac{w^{2}}{2}}\mathrm{d}w\label{eq:I1}
\end{align}

\noindent and

\begin{equation}
I_{2}=\frac{ABK_{i}e^{-r\left(T-t\right)}}{\sqrt{2\pi}}\intop_{-g_{2_{i}}}^{\infty}e^{-\frac{w^{2}}{2}}\mathrm{d}w
\end{equation}
\\

Now, we only need to solve $I_{1}$ and $I_{2}$ to find the value
of $c_{j}$ in terms of $S_{i}$. After that, we have:

\begin{align}
c_{j} & \thickapprox AB\left(S_{i}^{t}\right)^{\alpha\frac{\sigma_{j}}{\sigma_{i}}}e^{\left(\alpha\frac{\sigma_{j}}{\sigma_{i}}-1\right)\left(r+\frac{1}{2}\alpha\sigma_{j}\sigma_{i}\right)\left(T-t\right)}N\left(g_{1_{i}}\right)-ABK_{i}e^{-r\left(T-t\right)}N\left(g_{2_{i}}\right)
\end{align}

\noindent  where

\begin{eqnarray}
g_{1} & = & \frac{\ln\left(\frac{S_{i}^{t}}{K_{i}^{\sigma_{i}/\alpha\sigma_{j}}}\right)+\left[r+\left(\alpha\frac{\sigma_{j}}{\sigma_{i}}-\frac{1}{2}\right)\sigma_{i}^{2}\right]\left(T-t\right)}{\sigma_{i}\sqrt{T-t}}\\
 & = & g_{2}+\alpha\sigma_{j}\sqrt{T-t}
\end{eqnarray}
\\
Finally, and written in complete form, we have the option valuation
of asset $j$ using information of its twin asset $i$:

\begin{multline}
c_{j}\thickapprox S_{j}^{t}e^{\left[\left(\alpha\frac{\sigma_{j}}{\sigma_{i}}-1\right)\left(r+\frac{1}{2}\alpha\sigma_{j}\sigma_{i}\right)+\frac{1}{2}\sigma_{j}\left(\alpha\sigma_{i}-\sigma_{j}\right)\right]\left(T-t\right)}e^{\sigma_{j}\left(1-\rho\alpha\right)W_{x}^{T-t}-\alpha\sigma_{j}\sqrt{1-\rho^{2}}W_{y}^{T-t}}N\left(g_{1}\right)\\*
-S_{j}^{t}\left(S_{i}^{t}\right)^{-\alpha\frac{\sigma_{j}}{\sigma_{i}}}K_{i}e^{\left[\frac{1}{2}\sigma_{j}\left(\alpha\sigma_{i}-\sigma_{j}\right)-r\right]\left(T-t\right)}e^{\sigma_{j}\left(1-\rho\alpha\right)W_{x}^{T-t}-\alpha\sigma_{j}\sqrt{1-\rho^{2}}W_{y}^{T-t}}N\left(g_{2}\right)\\*
\label{eq:cjcomp-1}
\end{multline}
\\

From equation \ref{eq:cjcomp-1} we can see that in order to value
an option of the original asset called $j$, we need to have the following
parameters: $\mu_{j}$ , $\mu_{i}$ ,$\sigma_{j}$, $\sigma_{i}$,
and $\rho$. Besides we need the initial values of $S$. The $\alpha$
parameter will be implicit in the $\mu's$ and $\sigma's$. On the
other hand, the goodness of fit of the pricing will be given in terms
of the parameters $\rho$ and $\alpha$.

\pagebreak{}

\section{Numerical Illustration}

In this section, we develop a numerical illustration using the following
parameters $\mu_{i}=0.4$, $\mu_{j}=0.8$, $\sigma_{i}=0.2$, $S_{i}^{t=0}=80$
and $S_{j}^{t=0}=90$. Firstly, given these two assets we will model
one asset price path using its twin with different values of $\rho$
and $\alpha$, indicating its predictive power through its mean absolute
percentage error (MAPE). Secondly, using the same example we will
use our new option pricing formula and we will evaluate its performance
to predict a 3 month call option, again for different values of $\rho$
and $\alpha$.

\subsection{Twin Assets}

We simulate, under several values of $\rho$ and $\alpha$%
\footnote{The value of $\alpha$ was modified by changing $\mu_{j}$%
}, the path of the two stocks involved, $S_{i}$ and $S_{j}$ (Eqs.
\ref{eq:SiT} and \ref{eq:SjT}); where $S_{i}$ is the twin of $S_{j}$;
and compute, using \eqref{Sj_i compact}, the model prediction for
one day at future. These results are plotting in Fig. \ref{fig:Model-prediction}. 

We see in the Fig. \ref{fig:Model-prediction}, a perfect fit when
$\left(\rho,\alpha\right)=\left(1,1\right)$. We also see that if
we have a low value of correlation, the fit is least precise. On the
other hand, if $\alpha\neq1$ the results lose accuracy. In order
to have a more clear and systematic measure of error, we use a Montecarlo
simulation, with 40000 replications for each pair of values $\left(\rho,\alpha\right)$,
and then we compute the Mean Absolute Percentage Error (MAPE) of these
simulations. The MAPE was obtained using the following definition:

\begin{equation}
\mathrm{MAPE}\left(\rho_{l},\alpha_{m}\right)=\frac{100}{N}\sum_{n=1}^{N}\left|\frac{S'_{j_{n}}\left(\rho_{l},\alpha_{m}\right)-S{}_{j_{n}}}{S{}_{j_{n}}}\right|\label{eq:Mape}
\end{equation}

\noindent where $S'_{j_{n}}\left(\rho_{l},\alpha_{m}\right)$ is
the value of the $n^{th}$ simulation of the asset $S_{j}$ using
it twin (given by \ref{Sj_i est}) with parameters $\rho_{l}$ and
$\alpha_{m}$; $S{}_{j_{n}}$ is the $n^{th}$ simulation of the asset
$S_{j}$, and $N$ is the number of replications. The results of these
simulations are presented in \ref{fig:Mean-absolute-percentage}.
At the right side of the figure a fix value of $\alpha$ is shown;
in this case the MAPE is monotonically decreasing function (in relation
to$\rho$). In case of $\left(\rho,\alpha\right)=\left(1,1\right),$
we have a perfect twin (without error).

If we would estimate the error for another time at future, the MAPE
have the same form as above, but the values are obviously amplified
(see \ref{fig:Mean-absolute-percentage-1mes}).

Besides, the error (MAPE) is clearly related with the value of $\sigma_{j}$.
If the value of$\sigma_{j}$ increases, the error will grow. We can
see that behavior in fig.\ref{fig:Mapevolatilidades}. At the left
side of the image we have a $\rho=1$ and different values of $\alpha$;
and at the right side we fix $\alpha=1$ and different values of $\rho$.
We plot the MAPE for three different values of $\sigma_{j}$. In the
two subplots, the MAPE increases if $\sigma_{j}$ is greater.

\subsection{Option Pricing of Twin Assets}

Similar to the previous section, we shown in Fig. \ref{fig:Mape3m}
the MAPE of the proposed option pricing model. For this example we
use the same parameters of the above subsection, besides $K_{J}=S_{j}^{t=0}$
(exercise price) and $r=0.05$ (risk free rate). In the same way,
these errors were obtained using the same methodology than before:

\begin{equation}
\mathrm{MAPE2}\left(\rho_{l},\alpha_{m}\right)=\frac{100}{N}\sum_{n=1}^{N}\left|\frac{c'_{j_{n}}\left(\rho_{l},\alpha_{m}\right)-c{}_{j}}{c{}_{j}}\right|\label{eq:Mape-1}
\end{equation}

\noindent being $c{}_{j}$ the theoretical value of a call option
over the stock $S_{j}$ with strike price $K_{J}$ (see \ref{eq:BScj}),
$c'_{j_{n}}\left(\rho_{l},\alpha_{m}\right)$ the $n^{th}$ simulation
of $c_{j}$ using the relation \eqref{eq:cjcomp-1} with parameters
$\rho_{l}$ and $\alpha_{m}$; $S{}_{j_{n}}$ is the $n^{th}$ simulation
of the asset $S_{j}$, and $N=10000$. 

As might be expected, the minimum error occurred when both $\rho$
and $\alpha$ are close to one. In fact, only when $\rho=1$ and $\alpha\in\left[0.95,1.05\right]$the
MAPE was under 10\%. A MAPE under 40\% occurred when $\rho=1$ and
$\alpha\in\left[0.8,1.25\right]$. In general for values of $\left(\rho,\alpha\right)$
distant to $\left(1,1\right)$, the error value is very high, over
100\% for instance. As shown above, the MAPE depends on the level
of sigma, being proportional in this case to the value of$\sigma_{j}$.
In this particular case, negative and positive $\rho$ are not symmetric
in terms of error, when $\alpha$ is kept constant, in fact the error
is decreasing when $\rho$ increases.

\section{Conclusions}

How to price and hedge claims on nontraded assets are becoming increasingly
important matters. Indeed, cross hedging strategies are widely used
in option pricing practice today. Nevertheless, there are not many
analytical studies about the goodness of fit of these methods. 

In this paper, the concept of twin assets was introduced, focusing
the discussion precisely in what does it mean for two assets to be
similar. Our findings point to the fact that, in order to have very
similar assets, for example identical twins assets, high correlation
measures are not enough. Specifically, two basic criteria of similarity
are pointed out: i) the coefficient of variation of the assets and
ii) the correlation between assets, these parameters may be considered
as a measure of accuracy and precision respectively. 

An option pricing model of twin assets is also developed, allowing
us to price an option of one nontraded asset using information of
its twin asset, but this time knowing explicitly what theoretical
levels of errors we are facing. Our numerical illustrations show,
as might be expected, that the minimum error in option pricing occurred
when both correlation $(\rho)$ and the ratio of the coefficient of
variation $(\alpha)$ of the assets are close to one. 

The empirical calibration of the model for real situations is an interested
future research. Also a more realistic, but more complex analysis,
could consider the stochastic modelling of parameters $\rho$ and
$\alpha$.

\section*{References}

\bibliographystyle{plain}
\begin{btSect}{0C__Users_Achel_Desktop_13ene_OptionPricing}
\btPrintCited
\end{btSect}

\begin{center}
\begin{sidewaysfigure}
\noindent \begin{centering}
\includegraphics[width=20cm]{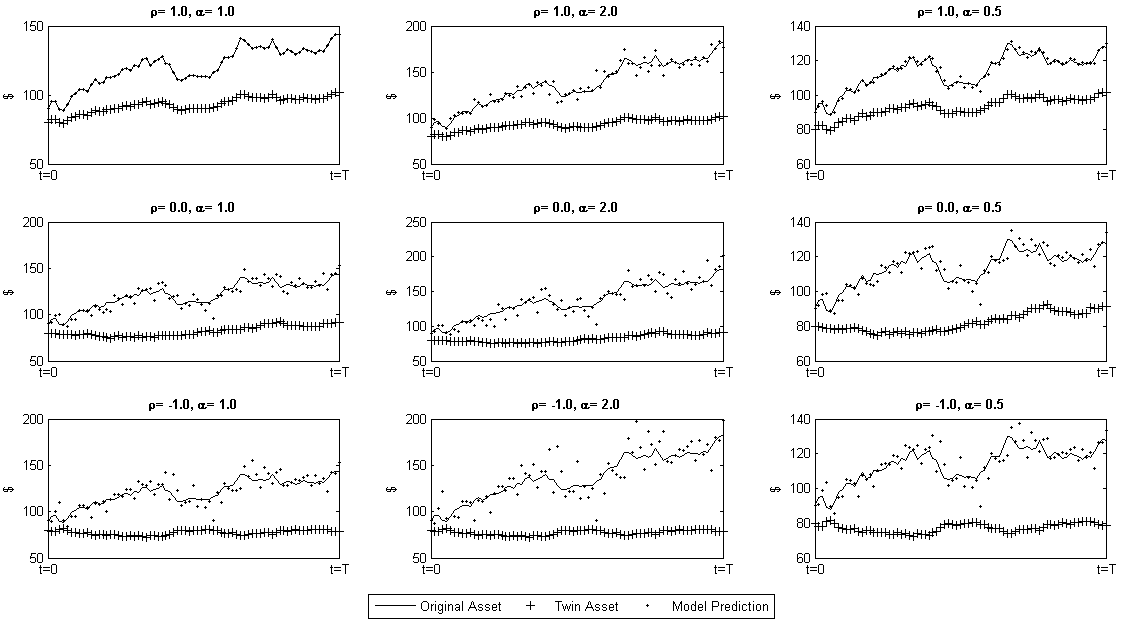}
\par\end{centering}

\caption{\label{fig:Model-prediction}Model prediction, for one day, of one
asset using it twin with different values of $\rho$ and $\alpha$.
We using $\mu_{i}=0.4$, $\mu_{j}=0.8$, $\sigma_{i}=0.2$, $S_{i}^{t=0}=80$
and $S_{j}^{t=0}=90$}
\end{sidewaysfigure}

\par\end{center}

\begin{center}
\begin{sidewaysfigure}
\noindent \begin{centering}
\includegraphics[width=20cm]{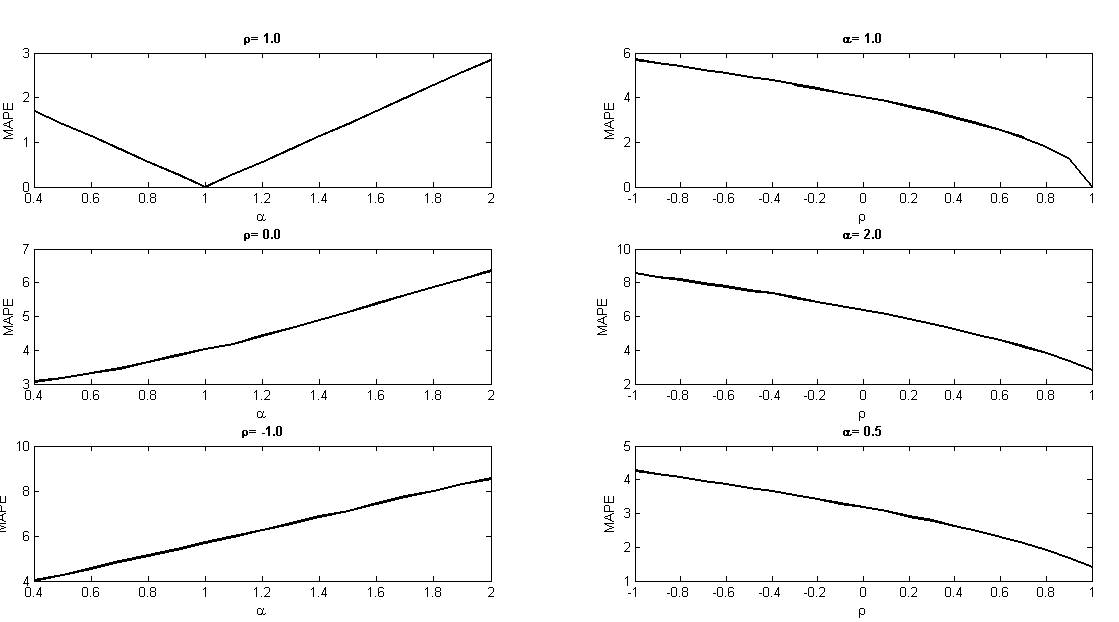}
\par\end{centering}

\caption{\label{fig:Mean-absolute-percentage}Mean absolute percentage error
for the Montecarlo model prediction, for one day, of one asset using
it twin.}
\end{sidewaysfigure}

\par\end{center}

\begin{center}
\begin{sidewaysfigure}
\noindent \begin{centering}
\includegraphics[width=20cm]{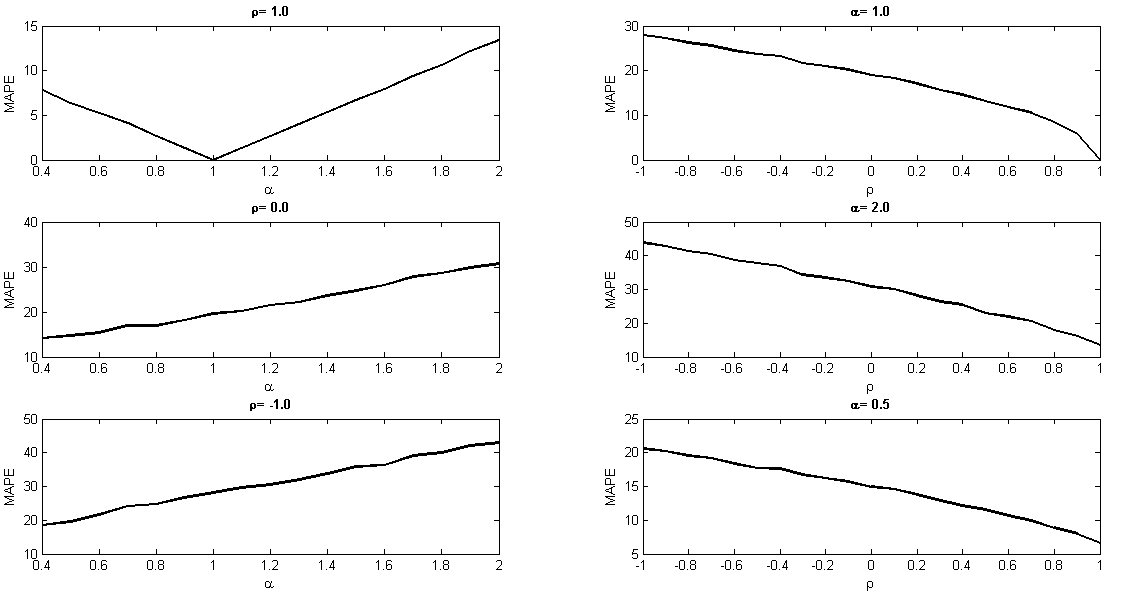}
\par\end{centering}

\caption{\label{fig:Mean-absolute-percentage-1mes}Mean absolute percentage
error for the Montecarlo model prediction, for one month, of one asset
using it twin.}
\end{sidewaysfigure}

\par\end{center}

\begin{center}
\begin{sidewaysfigure*}
\noindent \begin{centering}
\includegraphics[width=20cm]{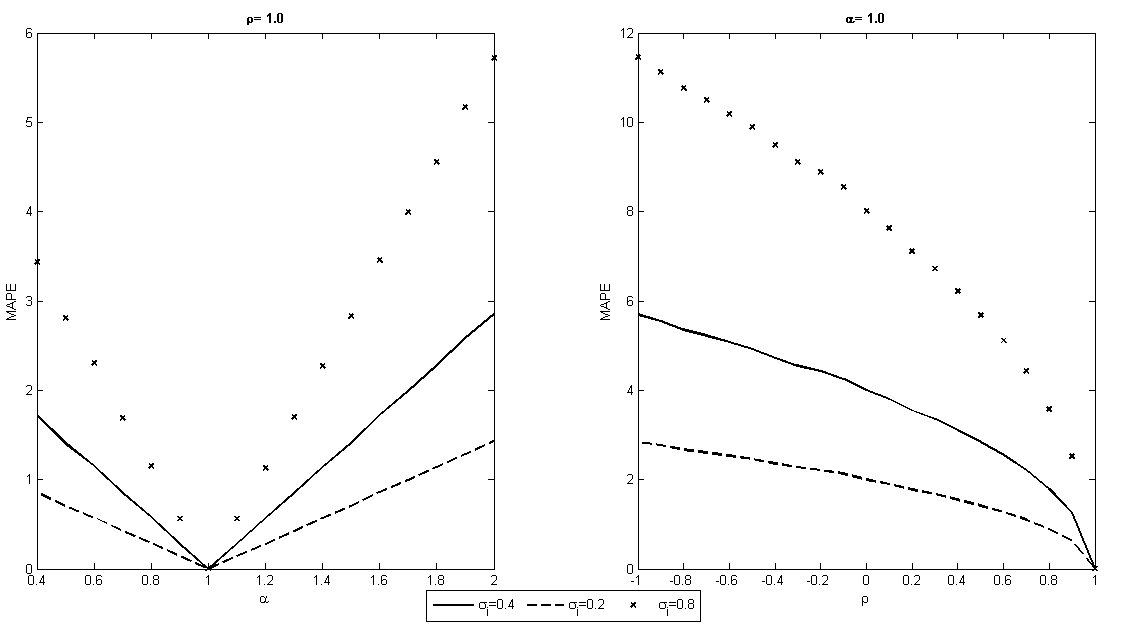}
\par\end{centering}

\caption{\label{fig:Mapevolatilidades}Mean absolute percentage error for the
Montecarlo model prediction, for one day, of one asset using it twin,
using different values of $\sigma_{j}$. }
\end{sidewaysfigure*}

\par\end{center}

\begin{center}
\begin{sidewaysfigure}
\noindent \begin{centering}
\includegraphics[width=20cm]{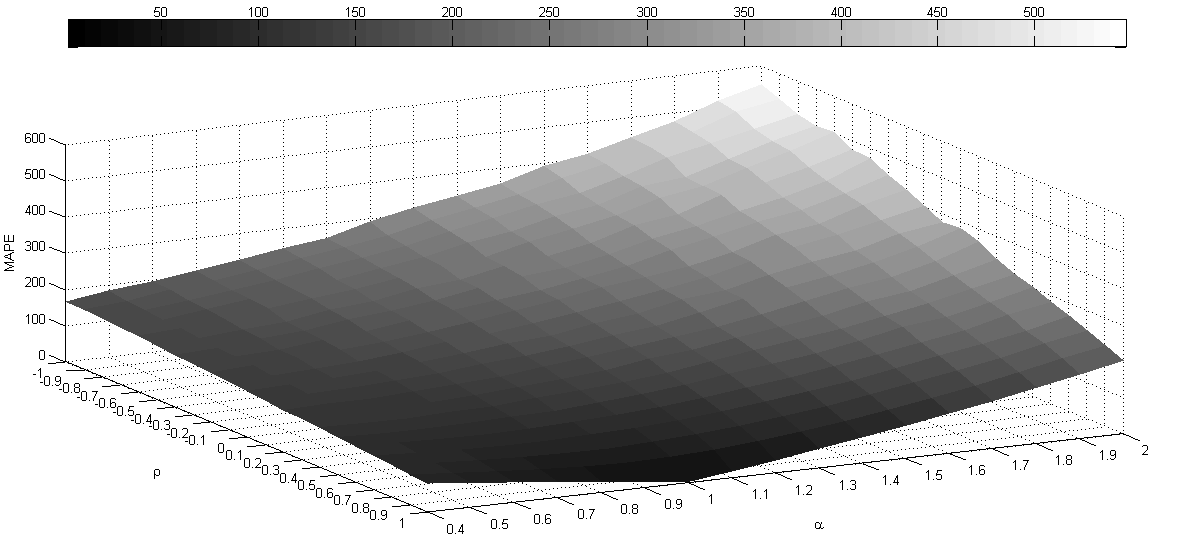}
\par\end{centering}

\caption{\label{fig:Mape3m}Mean absolute percentage error for the Montecarlo
model prediction, for a call option with three months of maturity,
using the twin asset approach.}
\end{sidewaysfigure}

\par\end{center}
\end{document}